\documentclass[graybox]{svmult}

\usepackage{graphicx}
\usepackage{epstopdf}
\usepackage{mathptmx}       % selects Times Roman as basic font
\usepackage{helvet}         % selects Helvetica as sans-serif font
\usepackage{courier}        % selects Courier as typewriter font
\usepackage{type1cm}        % activate if the above 3 fonts are
\usepackage{makeidx}         % allows index generation
\usepackage{graphicx}        % standard LaTeX graphics tool
\usepackage{multicol}        % used for the two-column index
\usepackage[bottom]{footmisc}% places footnotes at page bottom

\usepackage{latexsym,epsfig,graphicx,subfigure,amsmath,amssymb,amscd,multirow,psfrag,paralist,dsfont,float}

\makeindex             % used for the subject index
\newcommand{\thetavec}{{\boldsymbol{\theta}}}

\newcommand{\ti}{\mathrm{TI}}

\newcommand{\degreeC}{{^\circ\mathrm{C}}}

\makeatletter
\renewcommand\@biblabel[1]{#1.}
\makeatother

\begin{document}
%%%%%%%%%%%%%%%%%%%%%%%%%%%%%%%%%%%%%%%%%%%%%%%%%%%%%%%%%%%%%%%%%%%%%%%%%%%%%%%%%%%%%%%%%%%%%%%%
\title*{\texttt{ADDT}: An R Package for Analysis of Accelerated Destructive Degradation Test Data}
%\titlerunning{ADDT: An R Package for Analysis of Accelerated Destructive Degradation Test Data}
\author{Zhongnan Jin, Yimeng Xie, Yili Hong, and Jennifer H. Van~Mullekom}
\authorrunning{Jin, Xie, Hong, and Van~Mullekom}
\institute{Zhongnan Jin \at Department of Statistics, Virginia Tech, Blacksburg, VA 24061, \email{jinxx354@vt.edu}
\and Yimeng Xie \at Department of Statistics, Virginia Tech, Blacksburg, VA 24061, \email{xym@vt.edu}
\and Yili Hong,  corresponding author \at Department of Statistics, Virginia Tech, Blacksburg, VA 24061, \email{yilihong@vt.edu}
\and Jennifer H. Van~Mullekom \at Department of Statistics, Virginia Tech, Blacksburg, VA 24061, \email{vanmuljh@vt.edu}}

%%%%%%%%%%%%%%%%%%%%%%%%%%%%%%%%%%%%%%%%%%%%%%%%%%%%%%%%%%%%%%%%%%%%%%%%%%%%%%%%%%%%%%%%%
\maketitle
%%%%%%%%%%%%%%%%%%%%%%%%%%%%%%%%%%%%%%%%%%%%%%%%%%%%%%%%%%%%%%%%%%%%%%%%%%%%%%%%%%%%%%%%%
%online version abstract
\abstract*{Accelerated destructive degradation tests (ADDT) are often used to collect necessary data for assessing the long-term properties of polymeric materials. Based on the data, a thermal index (TI) is estimated. The TI can be useful for material rating and comparisons. The R package \texttt{ADDT} provides the functionalities of performing the traditional method based on the least-squares method, the parametric method based on maximum likelihood estimation, and the semiparametric method based on spline methods for analyzing ADDT data, and then estimating the TI for polymeric materials. In this chapter, we provide a detailed introduction to the \texttt{ADDT} package. We provide a step-by-step illustration for the use of functions in the package. Publicly available datasets are used for illustrations.}

%%%%%%%%%%%%%%%%%%%%%%%%%%%%%%%%%%%%%%%%%%%%%%%%%%%%%%%%%%%%%%%%%%%%%%%%%%%%%%%%%%%%%%%%%

%print version
\abstract{Accelerated destructive degradation tests (ADDT) are often used to collect necessary data for assessing the long-term properties of polymeric materials. Based on the data, a thermal index (TI) is estimated. The TI can be useful for material rating and comparisons. The R package \texttt{ADDT} provides the functionalities of performing the traditional method based on the least-squares method, the parametric method based on maximum likelihood estimation, and the semiparametric method based on spline methods for analyzing ADDT data, and then estimating the TI for polymeric materials. In this chapter, we provide a detailed introduction to the \texttt{ADDT} package. We provide a step-by-step illustration for the use of functions in the package. Publicly available datasets are used for illustrations.}

\newpage
%%%%%%%%%%%%%%%%%%%%%%%%%%%%%%%%%%%%%%%%%%%%%%%%%%%%%%%%%%%%%%%%%%%%%%%%%%%%%%%%%%%%%%%%%
\section{Introduction}\label{sec:introduction}
%%%%%%%%%%%%%%%%%%%%%%%%%%%%%%%%%%%%%%%%%%%%%%%%%%%%%%%%%%%%%%%%%%%%%%%%%%%%%%%%%%%%%%%%%
Accelerated destructive degradation tests (ADDT) are commonly used to collect data to access the long-term properties of polymeric materials (e.g., \cite{UL746B}). Based on the collected ADDT data, a thermal index (TI) is estimated using a statistical model. In practice, the TI can be useful for material rating and comparisons. In literature, there are three methods available for ADDT data modeling and analysis: the traditional method based on the least-squares approach, the parametric method based on maximum likelihood estimation, and the semiparametric method based on spline models. The chapter in Xie et al.~\cite{XieJinHongVanMullekom2017} provides a comprehensive review for the three methods for ADDT data analysis and compares the corresponding TI estimation procedures via simulations.

The R package \texttt{ADDT} in Hong et al.~\cite{Raddt} provides the functionalities of performing the three methods and their corresponding TI estimation procedures. In this chapter, we provide a detailed introduction to the \texttt{ADDT} package. We provide a step-by-step illustration for the use of functions in the package. We also use publicly available datasets for illustrations.

The rest of the chapter is organized as follows. Section~\ref{sec:R examples} introduces the three methods, the corresponding TI procedures, and the implementations in the R package. The Adhesive Bond B data (\cite{Escobaretal2003}) is used to do a step-by-step illustration. Section~\ref{sec:data.analysis} provides a full analysis of the Seal Strength data (\cite{LiDoganaksoy2014}) so that users can see a typical ADDT modeling and analysis process. Section~\ref{sec:concluding.remarks} contains some concluding remarks.

%%%%%%%%%%%%%%%%%%%%%%%%%%%%%%%%%%%%%%%%%%%%%%%%%%%%%%%%%%%%%%%%%%%%%%%%%%%%%%%%%%%%%%%%%%%%%%%%
\section{The Statistical Methods}\label{sec:R examples}
%%%%%%%%%%%%%%%%%%%%%%%%%%%%%%%%%%%%%%%%%%%%%%%%%%%%%%%%%%%%%%%%%%%%%%%%%%%%%%%%%%%%%%%%%%%%%%%%
\subsection{Data}
%%%%%%%%%%%%%%%%%%%%%%%%%%%%%%%%%%%%%%%%%%%%%%%%%%%%%%%%%%%%%%%%%%%%%%%%%%%%%%%%%%%%%%%%%%%%%%%%
In most applications, an ADDT dataset typically includes degradation measurements under different measuring time points, and accelerating variables such as temperature and voltage. In the \texttt{ADDT} package, there are four publicly available datasets ready for users to do analysis, which are the Adhesive Bond B data in~\cite{Escobaretal2003}, the Seal Strength data in~\cite{LiDoganaksoy2014}, the Polymer~Y data in~\cite{Tsaietal2013}, and the Adhesive Formulation K data in~\cite{XieKingHongYang2015}. Users can load those datasets by downloading, installing the package \texttt{ADDT} and appropriately calling the \emph{data} function. The following gives some example R codes.
\begin{verbatim}
>install.packages("ADDT")
>library(ADDT)
>data(AdhesiveBondB)
>data(SealStrength)
>data(PolymerY)
>data(AdhesiveFormulationK)
>AdhesiveBondB
>SealStength
\end{verbatim}

\begin{figure}
\begin{center}
\includegraphics[width=.8\textwidth]{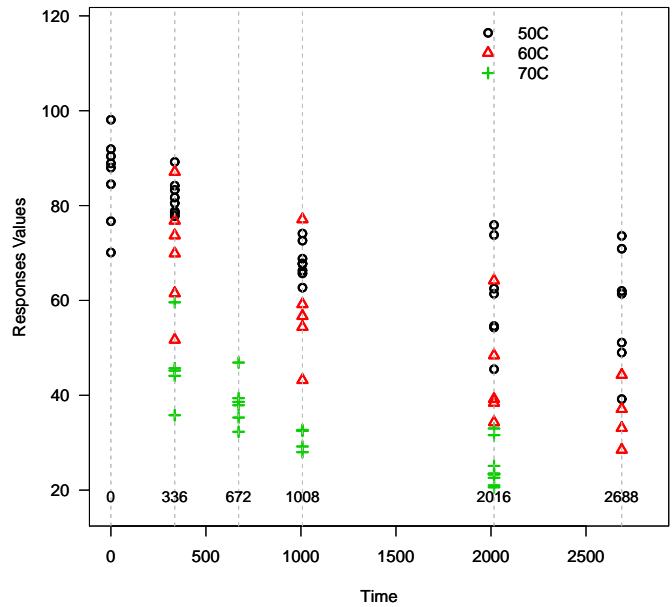}
\end{center}
\caption{Graphical representation of the Adhesive Bond B dataset. The x-axis  stands for the time in hour while y-axis represents the degradation values.}\label{plot: Adhesive Scatter plot}
\end{figure}

Table~\ref{tab:AdhesiveBondBData} shows the Adhesive Bond B dataset. The first column is the acceleration variable, temperature in Celsius. Time points that used to measure the degradation and the degradation values are listed in columns 2 and 3 correspondingly. We illustrate the Adhesive Bond B data in Fig \ref{plot: Adhesive Scatter plot}. To use the R \texttt{ADDT} package, users need to format the data in the same form as the dataset shown in Table~\ref{tab:AdhesiveBondBData}.

Another dataset that has been frequently used is the Seal Strength data where the strength from ten different seals were measured at five different time points under four different temperature levels. Seal Strength data is shown in Table~\ref{tab:SealStrengthData}. We will use the Adhesive Bond B data and Seal Strength data to illustrate the use of the \texttt{ADDT} package.

\begin{table}
\caption{The Adhesive Bond B data from Escobar et al.~\cite{Escobaretal2003}, which contains the testing of results of an ADDT for the strength of Adhesive Bond B.}\label{tab:AdhesiveBondBData}
\begin{center}
\begin{tabular}{ccc|ccc|ccc}\hline\hline
TempC & TimeH  & Response & TempC & TimeH  & Response & TempC & TimeH  & Response\\\hline
50 &    0 & 70.1 & 50 & 2016 & 62.5 &  60 & 2688 & 37.1 \\
50 &    0 & 76.7 & 50 & 2016 & 73.8 &  60 & 2688 & 44.3 \\
50 &    0 & 84.5 & 50 & 2016 & 75.9 &  70 &  336 & 35.8 \\
50 &    0 & 88.0 & 50 & 2688 & 39.2 &  70 &  336 & 44.1 \\
50 &    0 & 88.9 & 50 & 2688 & 49.0 &  70 &  336 & 45.2 \\
50 &    0 & 90.4 & 50 & 2688 & 51.1 &  70 &  336 & 45.7 \\
50 &    0 & 91.9 & 50 & 2688 & 61.4 &  70 &  336 & 59.6 \\
50 &    0 & 98.1 & 50 & 2688 & 62.0 &  70 &  672 & 32.3 \\
50 &  336 & 77.8 & 50 & 2688 & 70.9 &  70 &  672 & 35.3 \\
50 &  336 & 78.4 & 50 & 2688 & 73.6 &  70 &  672 & 37.9 \\
50 &  336 & 78.8 & 60 &  336 & 51.7 &  70 &  672 & 38.6 \\
50 &  336 & 80.5 & 60 &  336 & 61.5 &  70 &  672 & 39.4 \\
50 &  336 & 81.7 & 60 &  336 & 69.9 &  70 &  672 & 46.9 \\
50 &  336 & 83.3 & 60 &  336 & 73.7 &  70 & 1008 & 28.0 \\
50 &  336 & 84.2 & 60 &  336 & 76.8 &  70 & 1008 & 29.2 \\
50 &  336 & 89.2 & 60 &  336 & 87.1 &  70 & 1008 & 32.5 \\
50 & 1008 & 62.7 & 60 & 1008 & 43.2 &  70 & 1008 & 32.7 \\
50 & 1008 & 65.7 & 60 & 1008 & 54.4 &  70 & 2016 & 20.6 \\
50 & 1008 & 66.3 & 60 & 1008 & 56.7 &  70 & 2016 & 21.0 \\
50 & 1008 & 67.7 & 60 & 1008 & 59.2 &  70 & 2016 & 22.6 \\
50 & 1008 & 67.8 & 60 & 1008 & 77.1 &  70 & 2016 & 23.3 \\
50 & 1008 & 68.8 & 60 & 2016 & 34.3 &  70 & 2016 & 23.4 \\
50 & 1008 & 72.6 & 60 & 2016 & 38.4 &  70 & 2016 & 23.5 \\
50 & 1008 & 74.1 & 60 & 2016 & 39.2 &  70 & 2016 & 25.1 \\
50 & 2016 & 45.5 & 60 & 2016 & 48.4 &  70 & 2016 & 31.6 \\
50 & 2016 & 54.3 & 60 & 2016 & 64.2 &  70 & 2016 & 33.0 \\
50 & 2016 & 54.6 & 60 & 2688 & 28.5 &     &      &      \\
50 & 2016 & 61.4 & 60 & 2688 & 33.1 &     &      &      \\\hline\hline
\end{tabular}
\end{center}
\end{table}

\begin{table}
\caption{The Seal Strength data in Li and Daganaksoy~\cite{LiDoganaksoy2014}. The table shows the strength of seal samples that were measured at five different time points under four different temperature levels.}\label{tab:SealStrengthData}
\begin{center}
\begin{tabular}{ccc|ccc|ccc|ccc}\hline\hline
TempC & TimeH & Response & TempC & TimeH & Response & TempC  & TimeH & Response & TempC & TimeH & Response \\ \hline
100 & 0 & 28.74 & 200 & 1680 & 42.21 & 250 & 2520 & 17.08 & 300 & 3360 & 3.08 \\
100 & 0 & 25.59 & 200 & 1680 & 32.64 & 250 & 2520 & 11.52 & 350 & 3360 & 1.24 \\
100 & 0 & 22.72 & 200 & 1680 & 32.10 & 250 & 2520 & 13.03 & 350 & 3360 & 1.57 \\
100 & 0 & 22.44 & 200 & 1680 & 32.37 & 250 & 2520 & 18.37 & 350 & 3360 & 2.06 \\
100 & 0 & 29.48 & 200 & 1680 & 33.59 & 300 & 2520 & 3.86 & 350 & 3360 & 1.56 \\
100 & 0 & 23.85 & 200 & 1680 & 26.46 & 300 & 2520 & 4.76 & 350 & 3360 & 1.94 \\
100 & 0 & 20.24 & 200 & 1680 & 33.69 & 300 & 2520 & 5.32 & 350 & 3360 & 1.39 \\
100 & 0 & 22.33 & 250 & 1680 & 14.29 & 300 & 2520 & 3.74 & 350 & 3360 & 1.91 \\
100 & 0 & 21.70 & 250 & 1680 & 20.16 & 300 & 2520 & 4.58 & 350 & 3360 & 1.44 \\
100 & 0 & 27.97 & 250 & 1680 & 22.35 & 300 & 2520 & 3.62 & 350 & 3360 & 1.61 \\
200 & 840 & 52.52 & 250 & 1680 & 21.96 & 300 & 2520 & 3.58 & 350 & 3360 & 1.50 \\
200 & 840 & 30.23 & 250 & 1680 & 13.67 & 300 & 2520 & 3.47 & 200 & 4200 & 14.53 \\
200 & 840 & 31.90 & 250 & 1680 & 14.40 & 300 & 2520 & 3.29 & 200 & 4200 & 17.95 \\
200 & 840 & 33.15 & 250 & 1680 & 22.37 & 300 & 2520 & 3.63 & 200 & 4200 & 11.90 \\
200 & 840 & 34.26 & 250 & 1680 & 13.08 & 350 & 2520 & 1.34 & 200 & 4200 & 17.00 \\
200 & 840 & 31.82 & 250 & 1680 & 17.81 & 350 & 2520 & 0.92 & 200 & 4200 & 15.56 \\
200 & 840 & 27.10 & 250 & 1680 & 17.82 & 350 & 2520 & 1.31 & 200 & 4200 & 18.07 \\
200 & 840 & 30.00 & 300 & 1680 & 10.34 & 350 & 2520 & 1.76 & 200 & 4200 & 13.96 \\
200 & 840 & 26.96 & 300 & 1680 & 13.24 & 350 & 2520 & 1.30 & 200 & 4200 & 13.57 \\
200 & 840 & 42.73 & 300 & 1680 &  8.57 & 350 & 2520 & 1.47 & 200 & 4200 & 16.35 \\
250 & 840 & 28.97 & 300 & 1680 & 11.93 & 350 & 2520 & 1.11 & 200 & 4200 & 18.76 \\
250 & 840 & 35.01 & 300 & 1680 & 13.76 & 350 & 2520 & 1.25 & 250 & 4200 & 14.75 \\
250 & 840 & 27.39 & 300 & 1680 & 16.44 & 350 & 2520 & 1.02 & 250 & 4200 & 11.54 \\
250 & 840 & 36.66 & 300 & 1680 & 14.81 & 350 & 2520 & 1.30 & 250 & 4200 & 11.57 \\
250 & 840 & 27.91 & 300 & 1680 & 11.50 & 200 & 3360 & 26.72 & 250 & 4200 & 10.83 \\
250 & 840 & 31.03 & 300 & 1680 & 11.92 & 200 & 3360 & 21.24 & 250 & 4200 & 12.78 \\
250 & 840 & 32.65 & 300 & 1680 & 10.30 & 200 & 3360 & 22.76 & 250 & 4200 & 10.14 \\
250 & 840 & 35.08 & 350 & 1680 &  5.78 & 200 & 3360 & 24.39 & 250 & 4200 & 11.45 \\
250 & 840 & 28.05 & 350 & 1680 &  5.90 & 200 & 3360 & 15.93 & 250 & 4200 & 12.91 \\
250 & 840 & 33.54 & 350 & 1680 &  6.99 & 200 & 3360 & 23.90 & 250 & 4200 & 13.06 \\
300 & 840 & 10.63 & 350 & 1680 &  7.94 & 200 & 3360 & 22.09 & 250 & 4200 & 6.76 \\
300 & 840 &  8.28 & 350 & 1680 &  7.06 & 200 & 3360 & 23.69 & 300 & 4200 & 1.95 \\
300 & 840 & 13.46 & 350 & 1680 &  5.13 & 200 & 3360 & 23.67 & 300 & 4200 & 1.55 \\
300 & 840 & 13.47 & 350 & 1680 &  5.80 & 200 & 3360 & 20.94 & 300 & 4200 & 2.19 \\
300 & 840 &  9.44 & 350 & 1680 &  6.20 & 250 & 3360 & 14.23 & 300 & 4200 & 2.00 \\
300 & 840 &  7.66 & 350 & 1680 &  5.30 & 250 & 3360 & 12.83 & 300 & 4200 & 2.00 \\
300 & 840 & 11.16 & 350 & 1680 &  6.34 & 250 & 3360 & 13.02 & 300 & 4200 & 2.33 \\
300 & 840 &  8.70 & 200 & 2520 &  9.47 & 250 & 3360 & 16.74 & 300 & 4200 & 1.80 \\
300 & 840 &  9.44 & 200 & 2520 & 13.61 & 250 & 3360 & 12.11 & 300 & 4200 & 2.34 \\
300 & 840 & 12.23 & 200 & 2520 &  8.95 & 250 & 3360 & 12.24 & 300 & 4200 & 1.88 \\
350 & 840 & 13.79 & 200 & 2520 &  8.61 & 250 & 3360 & 18.97 & 300 & 4200 & 2.66 \\
350 & 840 & 15.10 & 200 & 2520 & 10.16 & 250 & 3360 & 15.29 & 350 & 4200 & 0.27 \\
350 & 840 & 20.58 & 200 & 2520 &  8.82 & 250 & 3360 & 14.38 & 350 & 4200 & 0.20 \\
350 & 840 & 18.20 & 200 & 2520 &  8.84 & 250 & 3360 & 14.80 & 350 & 4200 & 0.26 \\
350 & 840 & 16.64 & 200 & 2520 & 10.73 & 300 & 3360 & 2.89 & 350 & 4200 & 0.26 \\
350 & 840 & 10.93 & 200 & 2520 & 10.63 & 300 & 3360 & 3.31 & 350 & 4200 & 0.27 \\
350 & 840 & 12.28 & 200 & 2520 &  7.70 & 300 & 3360 & 1.81 & 350 & 4200 & 0.18 \\
350 & 840 & 18.65 & 250 & 2520 &  9.59 & 300 & 3360 & 1.61 & 350 & 4200 & 0.13 \\
350 & 840 & 20.80 & 250 & 2520 & 14.37 & 300 & 3360 & 2.65 & 350 & 4200 & 0.20 \\
350 & 840 & 15.04 & 250 & 2520 & 12.08 & 300 & 3360 & 2.83 & 350 & 4200 & 0.13 \\
200 & 1680 & 31.37 & 250 & 2520 & 11.79 & 300 & 3360 & 2.70 & 350 & 4200 & 0.21 \\
200 & 1680 & 37.91 & 250 & 2520 & 17.69 & 300 & 3360 & 2.79 &  &  & \\
200 & 1680 & 38.03 & 250 & 2520 & 14.05 & 300 & 3360 & 1.83 &  &  &  \\ \hline\hline
\end{tabular}
\end{center}
\end{table}

%%%%%%%%%%%%%%%%%%%%%%%%%%%%%%%%%%%%%%%%%%%%%%%%%%%%%%%%%%%%%%%%%%%%%%%%%%%%%%%%%%%%%%%%%%%%%%%%
\subsection{The Traditional Method}\label{sec:Least Square}
%%%%%%%%%%%%%%%%%%%%%%%%%%%%%%%%%%%%%%%%%%%%%%%%%%%%%%%%%%%%%%%%%%%%%%%%%%%%%%%%%%%%%%%%%%%%%%%%
The traditional method using the least-squares approach is widely accepted and used in various industrial applications. The traditional method is a two-step approach that uses polynomial fittings and the least-squares method to obtain the temperature-time relationship. The TI can be obtained by using the fitted temperature-time relationship. In particular, for each temperature level, indexed by $i$, we find the mean time to failure $m_i$ satisfies the following equation.
\begin{align}\label{eqn:tf.poly.intp}
a_{0i}+a_{1i}m_i+a_{2i}m_i^2+a_{3i}m_i^3=y_{f}, i=1,\cdots, n,
\end{align}
where $y_{f}$ is the failure threshold and $(a_{0i}, a_{1i}, a_{2i}, a_{3i})^{'}$ are coefficients. Here $n$ is the number of temperature levels. The temperature-time relationship is expressed as
\begin{align}\label{eqn:Arrihenius Relationship}
\log_{10}(m_i)=\beta_0+\beta_1x_i, i=1,\cdots, n,
\end{align}
which is based on the Arrhenius relationship to extrapolate to the normal use condition. With the parameterizations in this temperature-time relationship, the TI, denoted by R, can be estimated as:
\begin{align}\label{eqn:lsa.ti}
R=\frac{\beta_1}{\log_{10}(t_d)-\beta_0}-273.16\,.
\end{align}
where $\beta_{0}$ and $\beta_{1}$ are the same with the coefficients from equation \ref{eqn:Arrihenius Relationship}, and $t_{d}$ is the target time, usually $t_{d} = 100{,}000$ is used.

In the R package \texttt{ADDT}, we implement the traditional method by using:
\begin{verbatim}
>addt.fit.lsa<-addt.fit(Response~TimeH+TempC,data=Adh
esiveBondB, proc="LS", failure.threshold=70)
\end{verbatim}
The \emph{addt.fit} function in \texttt{ADDT} package fits the traditional model automatically when users specify \emph{proc = ``LS"} argument. In function \emph{addt.fit}, other arguments include:
\begin{itemize}
\item \emph{formula}: We use \emph{Response} $\sim$ \emph{TimeH+TempC} to represent the model formula. The \emph{Response}, \emph{TimeH}, and \emph{TempC} specify the response, time, and temperature columns in the dataset, respectively. Note that the order of \emph{TimeH} and \emph{TempC} can not be exchanged in the formula.

\item \emph{data}: The name of the dataset for analysis. The dataset should have the same layout as the Adhesive Bond B in Table~\ref{tab:AdhesiveBondBData}. Specifically, the order of the three columns should be the same as Adhesive Bond B, which is TempC, TimeH, and Response.

\item \emph{initial.value}: We need response measurements at time point 0 to compute the initial degradation level in the model. If the data do not contain that information, the user must supply the \emph{initial.value}. Otherwise, the function will give an error message.

\item \emph{failure.threshold}: This argument sets the failure threshold. The default value of the soft failure threshold is 70\% of the initial value in the \texttt{ADDT} package examples. Note that in industrial standards such as UL~746B~\cite{UL746B}, the failure threshold is usually 50\%.

\item \emph{time.rti}: The \emph{addt.fit} function allows users to specify the expected time associated with the TI. The default value for \emph{time.rti} is $t_{d}=$100{,}000 hours.

\item \emph{method}: This argument specifies the method that is used in the optimization process. Details can be found in \emph{optim} function in R. The default value is \emph{``Nelder-Mead''}.

\item \emph{subset}: This argument allows the users to specify a subset of the dataset for modeling.
\end{itemize}

The above arguments are the basic model inputs to run \emph{addt.fit},  when \emph{proc=``LS"}. Other methods, \emph{proc =``ML"} (the parametric method) and \emph{proc =``SemiPara"} (the semiparametric method) also require the same arguments. However, there are additional arguments for the other two methods and we will introduce them in Sections~\ref{sec:Maximum Likelihood} and~\ref{sec:Semiparametric}.

We store the model fitting results in the \emph{addt.fit.lsa} in this example. Users can print the model summary table and plots upon appropriate call. Examples are listed below:

\begin{center}
\begin{verbatim}
> summary(addt.fit.lsa)

Least Squares Approach:
    beta0     beta1
 -13.7805 5535.0907
est.TI: 22
Interpolation time:
     Temp      Time
[1,]   50 2063.0924
[2,]   60  797.1901
[3,]   70  206.1681
\end{verbatim}
\end{center}

\begin{figure}
\begin{center}
\includegraphics[width=.8\textwidth]{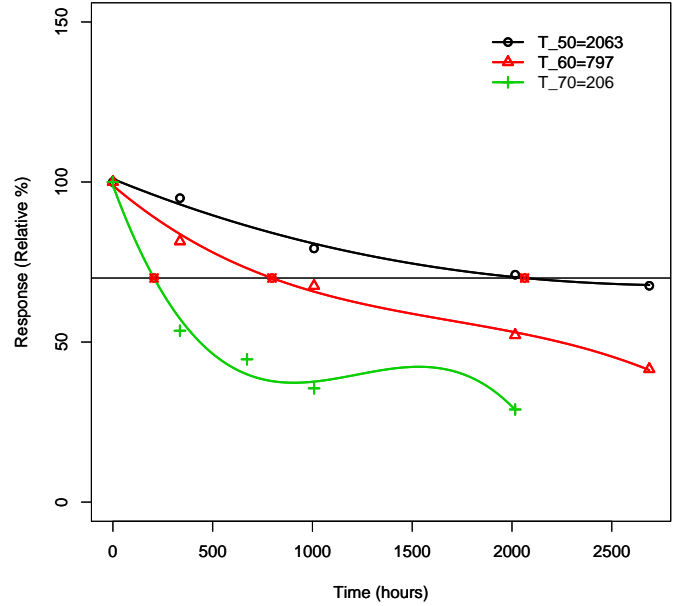}
\end{center}
\caption{Plot of the fitted polynomial curves for each temperature level, and the corresponding interpolated time to failures. The horizontal dark line presents the failure threshold. The y-axis shows the relative value of material strength.}\label{plot: LS plot}
\end{figure}

The \emph{summary} function for \emph{proc =``LS"} provides the parameter estimates and interpolated mean time to failure for the corresponding temperature levels. In the Adhesive Bond B example, the parameter estimates are $\hat{\beta}_{0} = -13.7805$ and  $\hat{\beta}_{1} = 5535.0907$ for the temperature-time relationship. Estimated mean time to failure for temperature level $50\degreeC$, $60\degreeC$, and $70\degreeC$, are 2063.092, 797.190 and 206.168 hours, respectively. The estimated TI is $22\degreeC$ in this example. Figure~\ref{plot: LS plot} shows the fitted polynomial curves for each temperature levels and the corresponding interpolated mean time to failure, according to least-squares method. The R code that is used to plot the results is shown below.
\begin{verbatim}
>plot(addt.fit.lsa, type="LS")
\end{verbatim}

%%%%%%%%%%%%%%%%%%%%%%%%%%%%%%%%%%%%%%%%%%%%%%%%%%%%%%%%%%%%%%%%%%%%%%%%%%%%%%%%%%%%%%%%%%%%%%%%
\subsection{The Parametric Method}\label{sec:Maximum Likelihood}
%%%%%%%%%%%%%%%%%%%%%%%%%%%%%%%%%%%%%%%%%%%%%%%%%%%%%%%%%%%%%%%%%%%%%%%%%%%%%%%%%%%%%%%%%%%%%%%%
Different from the two-step approach in the traditional method, for the parametric method, one uses a parametric model to describe the degradation path. The maximum likelihood (ML) method is then used to estimate the unknown parameters in the model. In particular, we assume that degradation measurement $y_{ijk}$ at time $t_{ij}$ for temperature level $i$ follows the model:
\begin{align*}
&y_{ijk} = \mu(t_{ij}; x_{i}) + \epsilon_{ijk}, i=1,\cdots,n, j=1,\cdots, n_{i}, k=1,\cdots, n_{ij},
\end{align*}
where
\begin{align*}
&x_{i}= \frac{1}{\textrm{TempC}_{i}+273.16},
\end{align*}
$\textrm{TempC}_{i}$ is the temperature level, the value $273.16$ is used to convert the temperature to Kelvin temperature scale. Here $n$ is the number of temperature levels, $n_i$ is the number of time points for level $i$, and $n_{ij}$ is the number samples tested under the temperature time combination and $\epsilon_{ijk}$ is the error term. For polymer materials, the following parametric assumption for $\mu(t; x)$ (e.g., \cite{VacaTrigoMeeker2009}) is used
\begin{align}\label{EQ: ParametricArr}
&\mu(t; x) = \frac{\alpha}{1 + [\frac{t}{\eta(x)} ]^{\gamma}},
\end{align}
where $\alpha$ represents the initial degradation, and $\gamma$ is a shape parameter. Here,
\begin{align*}
&\eta(x) = \exp(\upsilon_{0} + \upsilon_{1} x).
\end{align*}
is the scale factor that is based on the Arrhenius relationship. By the parametric specification, the ML method is then used to estimate the parameters. King et al.~\cite{Kingetal2016} performed a comprehensive comparison between the traditional method and the parametric method. Xie et al.~\cite{XieJinHongVanMullekom2017} performed a comprehensive comparison among the three methods in term of TI estimation.

For the model in \eqref{EQ: ParametricArr}, the TI is calculated as follows:
\begin{align}\label{eqn:lsa.ti.mle}
\ti=\frac{\beta_1}{\log_{10}(t_d)-\beta_0}-273.16,
\end{align}
where $\beta_{0}$ and $\beta_{1}$ are defined as:
$$\beta_0=\frac{\nu_0}{\log(10)}+\frac{1}{\gamma\log(10)}\log\left[\frac{1-p}{p}\right], \quad \textrm{and}\quad \beta_1=\frac{\nu_1}{\log(10)}.$$

To fit the parametric model, one can use the following command:
\begin{verbatim}
> addt.fit.mla<-addt.fit(Response~TimeH+TempC,data=Adh
esiveBondB,proc="ML", failure.threshold=70)
\end{verbatim}

Similar to the ``LS'' case, here we provide an example of \emph{ML} method based on the parametric method implemented in R. Using the same dataset Adhesive Bond B, we now change the \emph{proc} argument to \emph{proc =``ML"} so that the parametric model is used. The model results are stored in \emph{addt.fit.mla}. Argument setups are almost the same as those in \emph{addt.fit} for the case of \emph{proc = ``LS"} except for additional arguments: \emph{``starts"} and \emph{``fail.thres.vec"}. In particular,
\begin{itemize}
\item \emph{starts}: It provides a set of starting values for the ML estimation procedure. If this value is not supplied, the function will use the least-squares method to estimate for a set of starting values for the ML estimation.

\item \emph{fail.thres.vec}: If the user does not specify \emph{starts} argument, the user may instead provide a vector of two different \emph{failure.thresholds}. The least-squares procedure is then used for the two different failure thresholds to produce starting values for the ML procedure.

\end{itemize}

For the model results in \emph{addt.fit.lma}, we not only have the parameter estimates as in the \emph{LS} example, but also have confidence intervals for the model parameters and the TI. The following shows the summary information of the model fitting.
\begin{verbatim}
> summary(addt.fit.mla)

 Maximum Likelihood Approach:
Call:
lifetime.mle(dat = dat0, minusloglik = minus.loglik.ki
netics,  starts = starts, method = method, control =
 list(maxit = 1e+05))

Parameters:
            mean       std  95% Lower  95% Upper
alpha    87.2004    2.5920    82.2653    92.4315
beta0   -37.2360    4.6450   -46.3401   -28.1318
beta1 14913.1628 1561.1425 11853.3235 17973.0022
gamma     0.7274    0.0870     0.5753     0.9195
sigma     8.2017    0.6405     7.0377     9.5581
rho       0.0000    0.0003    -0.0006     0.0006

Temperature-Time Relationship:
     beta0      beta1
 -16.6830   6478.5641

TI:
      est       std 95% Lower 95% Upper
  25.6183    3.0980   19.5465   31.6902

Loglikelihod:
[1] -288.9057
\end{verbatim}
By applying \emph{summary} function to the \emph{addt.fit} results, we have the ML estimates for $\alpha, \upsilon_{0}, \upsilon_{1}, \gamma, \sigma$, and $\rho$ along with their standard deviation as well as the associated 95\% confidence intervals based on large-sample approximations. The log likelihood values for the final model is also printed for model comparisons.

The summary table will perform the TI estimates and confidence interval calculation automatically by assigning the default confidence level as 95\%. Users can change the confidence level to other values by using the function \emph{addt.confint.ti.mle} and specifying the desired value for \emph{conflevel}. In particular,
\begin{verbatim}
> addt.confint.ti.mle(addt.fit.mla, conflevel = 0.99)
\end{verbatim}
provides an example of customizing confidence level for TI estimates. It shows that the 99\% confidence interval for TI and the confidence interval is wider than using 95\% as the confidence level. The results are shown as follows.
\begin{verbatim}
     est.      s.e.     lower     upper
  25.618     3.097     17.638    33.598
\end{verbatim}

\begin{figure}
\begin{center}
\includegraphics[width=.8\textwidth]{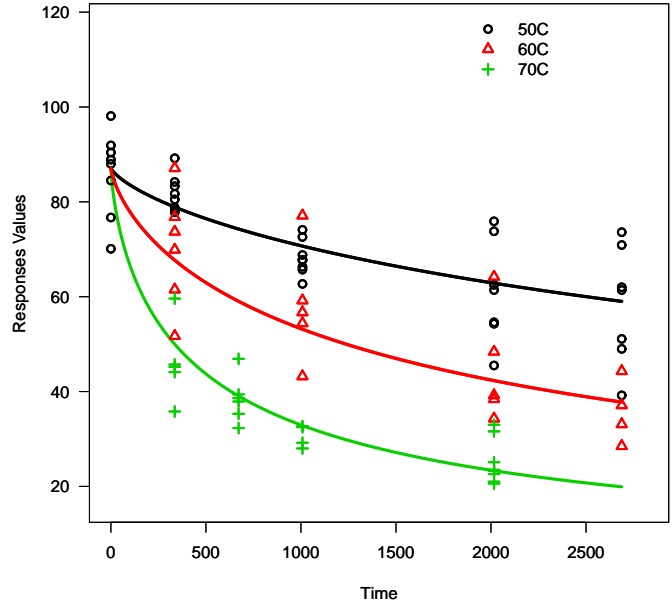}
\end{center}
\caption{Plot of the original dataset of Adhesive Bond B as well as the fitted degradation paths based on the parametric model. The black line, red line and green line stand for fitted lines at 50, 60 and 70 degree, respectively.}\label{plot: Parametric}
\end{figure}

Similar to the \emph{LS} method, we can visualize the model fitting results. For the \emph{ML} method, one can plot the fitted lines along with the data by employing \emph{plot.addt.fit}. Figure \ref{plot: Parametric} shows the illustration of the fitting results of \emph{plot.addt.fit}.

\begin{verbatim}
> plot(addt.fit.mla, type="ML")
\end{verbatim}

%%%%%%%%%%%%%%%%%%%%%%%%%%%%%%%%%%%%%%%%%%%%%%%%%%%%%%%%%%%%%%%%%%%%%%%%%%%%%%%%%%%%%%%%%%%%%%%%
\subsection{The Semiparametric Method}\label{sec:Semiparametric}
%%%%%%%%%%%%%%%%%%%%%%%%%%%%%%%%%%%%%%%%%%%%%%%%%%%%%%%%%%%%%%%%%%%%%%%%%%%%%%%%%%%%%%%%%%%%%%%%
Different from the traditional method and the parametric method that are introduced in Sections~\ref{sec:Least Square} and \ref{sec:Maximum Likelihood}, the semiparametric method is applicable to different materials with a nonparametric form for the baseline degradation path. In addition, the parametric part of the model (i.e., the Arrhenius relationship) retains the extrapolation capacity to the use condition. Similarly to the parametric model, we model the degradation measurement as follows,
$$y_{ijk} = \mu(t_{ij}, x_{i}; \theta) + \epsilon_{ijk},$$
where
$$x_{i}= -\frac{11605}{\textrm{TempC}_{i}+273.16},$$
and $\thetavec$ stands for all the parameters in the model. We use the semiparametric model structure to describe the degradation path. In particular, the degradation path is modeled as
\begin{align}\label{DegradationEquation}
&\mu(t_{ij}, x_{i}) = g[ \eta_{i}(t_{ij}; \beta) ; \gamma],
\end{align}
and the scale factor is
\begin{align}\label{DegradationTemperature}
&\eta_{i}(t; \beta) = \frac{t}{\exp(\beta s_{i})},
\end{align}
with acceleration parameter $\beta$. In equation \ref{DegradationTemperature}, we define $s_{i} = x_{max} - x_{i}$ where $x_{max}$ is the transformed value of the highest level of temperature.  We assume the error terms follow normal distribution with variance $\sigma^2$ and the correlations between two error terms are $\rho$. That is,
\begin{align*}
\epsilon_{ijk} \sim \textrm{N}(0, \sigma^2),
\end{align*}
and
\begin{align}\label{Equ: error term}
\textrm{Corr}(\epsilon_{ijk} , \epsilon_{ijk'}) = \rho.
\end{align} We assume $k \neq k'$ in the error terms correlations in \eqref{Equ: error term}. In \eqref{DegradationEquation}, $g(\cdot)$ is a monotonically decreasing function modeled by splines with parameter vector $\gamma$. See Xie et al.~\cite{XieKingHongYang2015} for more details on the semiparametric method.

As a more flexible method designated to a wide variety of materials, the non-parametric component is used to build the baseline degradation path. With inner knots $d_{1} \leq d_{2} \leq \cdots \leq d_{N}$ and boundary knots $d_{0}$, $d_{N+1}$, the $l$-th B-spline basis function with a degree of $q$ can be expressed at $z$ by recursively building the following models:
\begin{align*}
B_{0,l}(z) &= 1(d_{l} \leq z \leq d_{l+1})B_{q,l}(z) \\
&= \frac{z - d_{l}}{d_{l+q} - d_{l}} B_{q-1, l}(z) + \frac{ d_{l+q+q} - z}{d_{l+q+1} - d_{l+1}} B_{q-1, l+1}(z).
\end{align*}
The degradation can be expressed as follows.
\begin{align*}
&y_{ijk} = \sum_{l=1}^{p} \gamma_{l} B_{q,l}[\eta_{i}(t_{ij};\beta)] + \epsilon_{ijk},
\end{align*}
where $\eta(t;\beta)$ accounts for the parametric part while $g(\cdot)$ is the non-parametric component which is constrained to be monotonically decreasing to retain the meanings of the degradation process.

Similarly to the ``LS'' and ``ML'' methods, we implement the semiparametric model in R. In \emph{addt.fit}, \emph{proc=``SemiPara"} enables users to fit a semiparametric model to the degradation data as we discussed above. In particular,
\begin{verbatim}
>addt.fit.semi<-addt.fit(Response~TimeH+TempC,data=Adh
esiveBondB,proc="SemiPara",failure.threshold=70)
\end{verbatim}
Other than the arguments we introduced for \emph{proc = ``LS"} and \emph{proc = ``ML"}, there is an other unique option in the \emph{addt.fit} when \emph{proc = ``SemiPara"} is called. That is:
\begin{itemize}
\item \emph{semi.control}: This argument contains a list of control parameters regarding the \emph{SemiPara} option. Users can specify the model assumptions like correlation \emph{rho}. In \emph{semi.control = list(cor = F, \ldots)}, the default value is to exclude the correlation term in the model (i.e., $\rho=0$). If \emph{cor = T}, then there will be a correlation term in the semiparametric model.

\end{itemize}
Summary results of the semiparametric model object given by \emph{addt.fit} include $\hat{\beta}$, $\hat{\rho}$, knots that were used by the model, log-likelihood and AICc for the final model, which are both model evaluation quantities. Note that in the example shown below, we use the default set up for semiparametric model fit on the Adhesive Bond B data.

\begin{verbatim}
> summary(addt.fit.semi)

Semi-Parametric Approach:

Parameters Estimates:
betahat
  1.329

TI estimates:
 TI.semi    beta0    beta1
  26.313  -17.363 6697.074

Model Evaluations:
Loglikelihood      AICC
     -288.135   586.269

B-spline:
Left Boundary   knots   Right Boundary
        0.00   180.66         2016.00
\end{verbatim}

\begin{figure}
\begin{center}
\includegraphics[width=.8\textwidth]{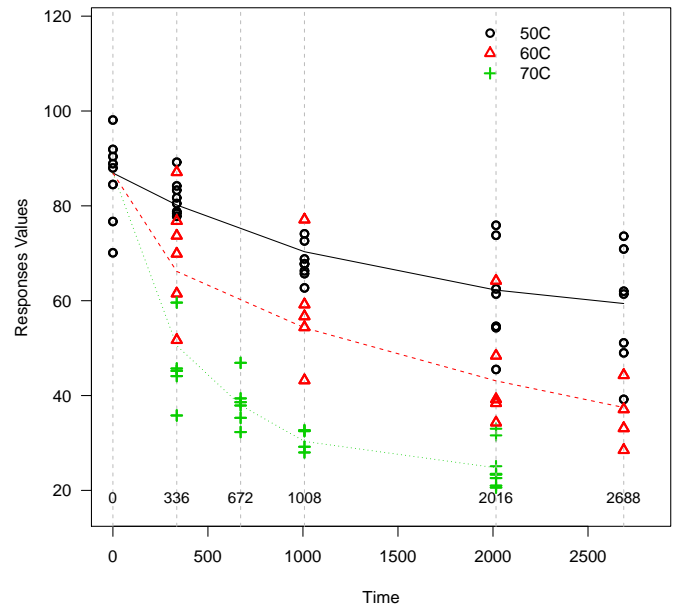}
\end{center}
\caption{Plot of the original dataset of Adhesive Bond B data as well as the fitted degradation mean values using the semiparametric model.}
\label{plot: SemiPara}
\end{figure}

We can also call \emph{plot.addt.fit} to present model fitting results.
\begin{verbatim}
plot(addt.fit.semi, type="SEMI")
\end{verbatim}
Figure \ref{plot: SemiPara} shows the plot of the original dataset of Adhesive Bond B data as well as the fitted degradation mean values using the semiparametric model. Here we assume that there is no correlation $\rho$ between two error terms. Note that for \emph{plot.addt} function, \emph{type} argument should be compatible with the \emph{addt.obj}, meaning that type used in plot function should be the same with \emph{proc} argument in the function \emph{addt.fit}, otherwise error messages will be generated.

\begin{figure}
\begin{center}
\includegraphics[width=.8\textwidth]{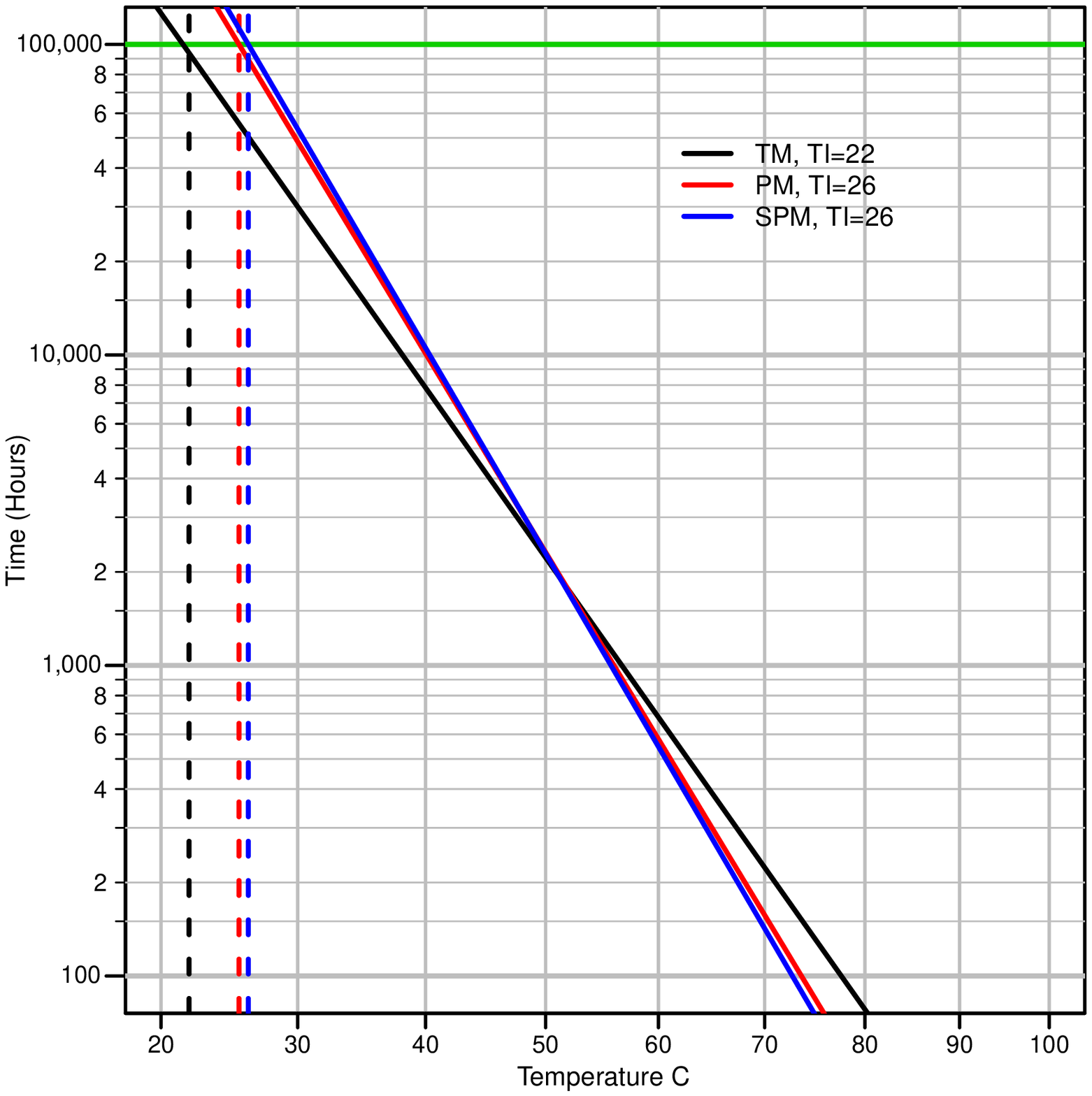}
\end{center}
\caption{Fitted temperature-time relationship lines for the Adhesive Bond B data from the least-squares, maximum likelihood, and the semiparametric methods. The failure threshold of 70\%.}
\label{plot: AdhesiveBondB TI comparison}
\end{figure}

We illustrate the comparisons among the least-squares, maximum likelihood and semiparametric methods in terms of TI estimation in Figure \ref{plot: AdhesiveBondB TI comparison}. Temperature-time relationship lines are plotted for all three methods in black, red and blue lines correspondingly.

%%%%%%%%%%%%%%%%%%%%%%%%%%%%%%%%%%%%%%%%%%%%%%%%%%%%%%%%%%%%%%%%%%%%%%%%%%%%%%%%%%%%%%%%%%%%%%%%
\section{Data Analysis}\label{sec:data.analysis}
%%%%%%%%%%%%%%%%%%%%%%%%%%%%%%%%%%%%%%%%%%%%%%%%%%%%%%%%%%%%%%%%%%%%%%%%%%%%%%%%%%%%%%%%%%%%%%%%
In this section, we present a complete ADDT data analysis using the Seal Strength data to illustrate the use of functions in Section~\ref{sec:R examples}. The details of the Seal Strength data is available in Li and Doganaksoy~\cite{LiDoganaksoy2014}. The first ten observations are listed below. Note that in the Seal Strength data, temperatures at time point 0 are modified to 200 degrees while those in the original Seal Strength dataset in Table \ref{tab:SealStrengthData} are 100 degrees. Changing temperatures at time point 0 to the lowest temperature is a computing trick that will not affect fitting results, because at time 0, the temperature effect has not kicked in yet.
\begin{center}
\begin{verbatim}
>head(SealStrength, n=10)
    TempC TimeH Response
1    200     0    28.74
2    200     0    25.59
3    200     0    22.72
4    200     0    22.44
5    200     0    29.48
6    200     0    23.85
7    200     0    20.24
8    200     0    22.33
9    200     0    21.70
10   200     0    27.97
\end{verbatim}
\end{center}

A graphical representation of the data is useful for users to obtain a general idea of the degradation paths. Using the \emph{addt.fit.mla} object from \emph{addt.fit} with \emph{proc=``ML"}, one can plot the degradation paths using option \emph{type=``data"}.
\begin{verbatim}
>plot(addt.fit.mla, type="data")
\end{verbatim}

\begin{figure}
\begin{center}
\includegraphics[width=.8\textwidth]{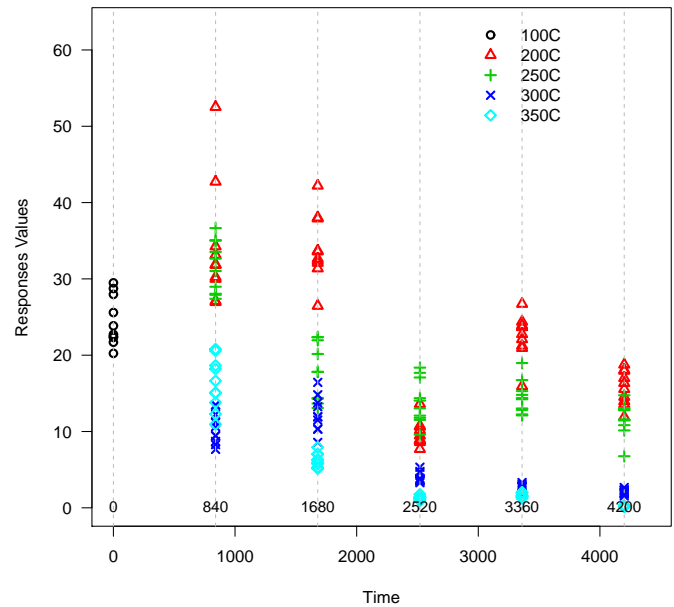}
\end{center}
\caption{Plot of the Seal Strength data. Degradations were measured at six different time points under three different temperatures.}\label{fig:seal.strength}
\end{figure}

Figure~\ref{fig:seal.strength} shows the plot of the Seal Strength data, in which the degradations were measured at six different time points under three different temperatures. For Seal Strength data, we observe an average decrease in degradation measurements as time increases. Degradation measurements decrease with the accelerating variable, temperature as well.

Three different \emph{addt.fit} models can be fitted, which are \emph{proc =``LS"}, \emph{proc = ``ML"}, and \emph{proc= ``SemiPara"}.
\begin{verbatim}
>addt.fit.lsa<-addt.fit(Response~TimeH+TempC,data=Seal
Strength,proc="LS",failure.threshold=70)

>addt.fit.mla<-addt.fit(Response~TimeH+TempC,data=Seal
Strength,proc="ML",failure.threshold=70)

>addt.fit.semi<-addt.fit(Response~TimeH+TempC,data=Seal
Strength,proc="SemiPara",failure.threshold=70)
\end{verbatim}

Alteratively, users can specify all three methods via one call of \emph{addt.fit} by setting \emph{proc = ``All"}. The returned object for the three methods is stored in \emph{addt.fit.all}.
\begin{verbatim}
> addt.fit.all<-addt.fit(Response~TimeH+TempC,data=Seal
Strength,proc="All",=failure.threshold=70)
\end{verbatim}

To view the results of all three models, users can call the \emph{summary} function:
\begin{verbatim}
> summary(addt.fit.all)
Least Squares Approach:
    beta0     beta1
   0.1934 1565.1731
est.TI: 52
Interpolation time:
     Temp      Time
[1,]  200 2862.3430
[2,]  250 2282.3303
[3,]  300  509.2084
[4,]  350  622.0857

 Maximum Likelihood Approach:
Call:
lifetime.mle(dat = dat0, minusloglik = minus.
loglik.kinetics, starts = starts, method =
method, control = list(maxit = 1e+05))

Parameters:
           mean      std 95% Lower 95% Upper
alpha   30.5898   3.4550   24.5152   38.1697
beta0    0.2991   1.7013   -3.0355    3.6337
beta1 3867.7170 899.5312 2104.6360 5630.7981
gamma    1.6556   0.4171    1.0105    2.7127
sigma    5.5456   0.6521    4.4041    6.9831
rho      0.7306   0.0664    0.6004    0.8607

Temperature-Time Relationship:
    beta0     beta1
  -0.0942 1680.4055

TI:
      est       std 95% Lower 95% Upper
  56.6920   28.1598    1.4997  111.8842

Loglikelihood:
[1] -555.0169

 Semi-Parametric Approach:

Parameters Estimates:
betahat
  0.282

TI estimates:
 TI.semi    beta0    beta1
  32.768    0.362 1418.833

Model Evaluations:
Loglikelihood          AICC
     -639.206      1288.412

B-spline:
 Left Boundary  knots  knots  knots   knots
          0.00 268.60 527.17 840.00 1394.55
 Right Boundary
 4200.00
\end{verbatim}

Results shown here are the same when users call \emph{summary} for three different models separately. The \emph{add.fit.all} and \emph{summary} for \emph{addt.fit.all} provides an alternative way to analyze the data simultaneously.

\begin{figure}
\begin{center}
\includegraphics[width=.8\textwidth]{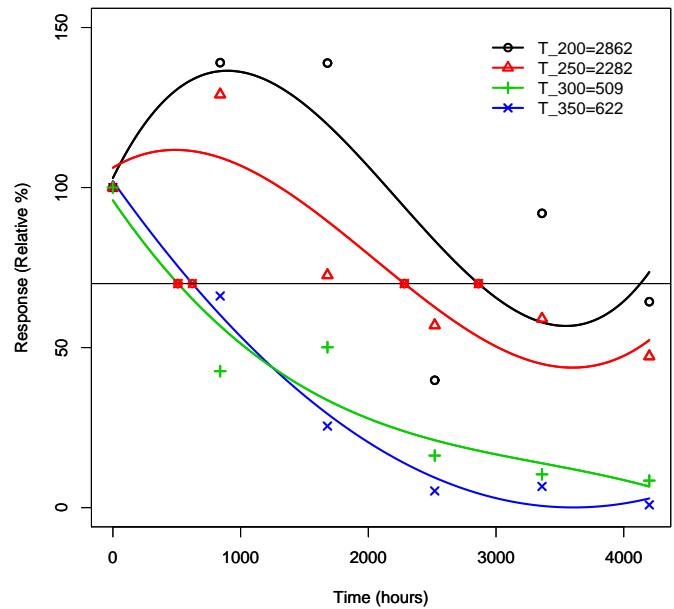}
\end{center}
\caption{Plot of the Seal Strength parametric lines with the least-squares method. The red, green, blue, light blue lines represent 200, 250, 300 and 350 degrees Celsius interpolated curves, respectively.}
\label{plot: Seal Strength LS}
\end{figure}

\begin{figure}
\begin{center}
\includegraphics[width=.8\textwidth]{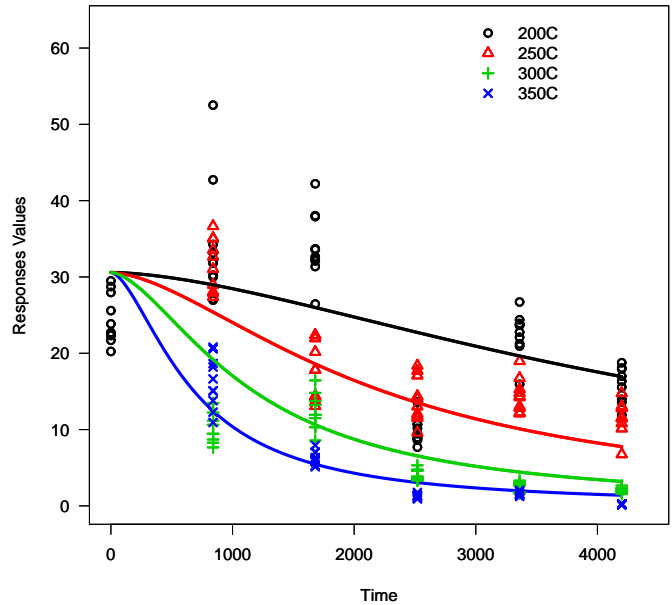}
\end{center}
\caption{Plot of the fitted mean function using maximum likelihood method for the Seal Strength data. The 200, 250, 300 and 350 degrees Celsius estimated curves are represented  by red, green, blue and light blue lines, respectively.}
\label{plot: Seal Strength ML}
\end{figure}

\begin{figure}
\begin{center}
\includegraphics[width=.8\textwidth]{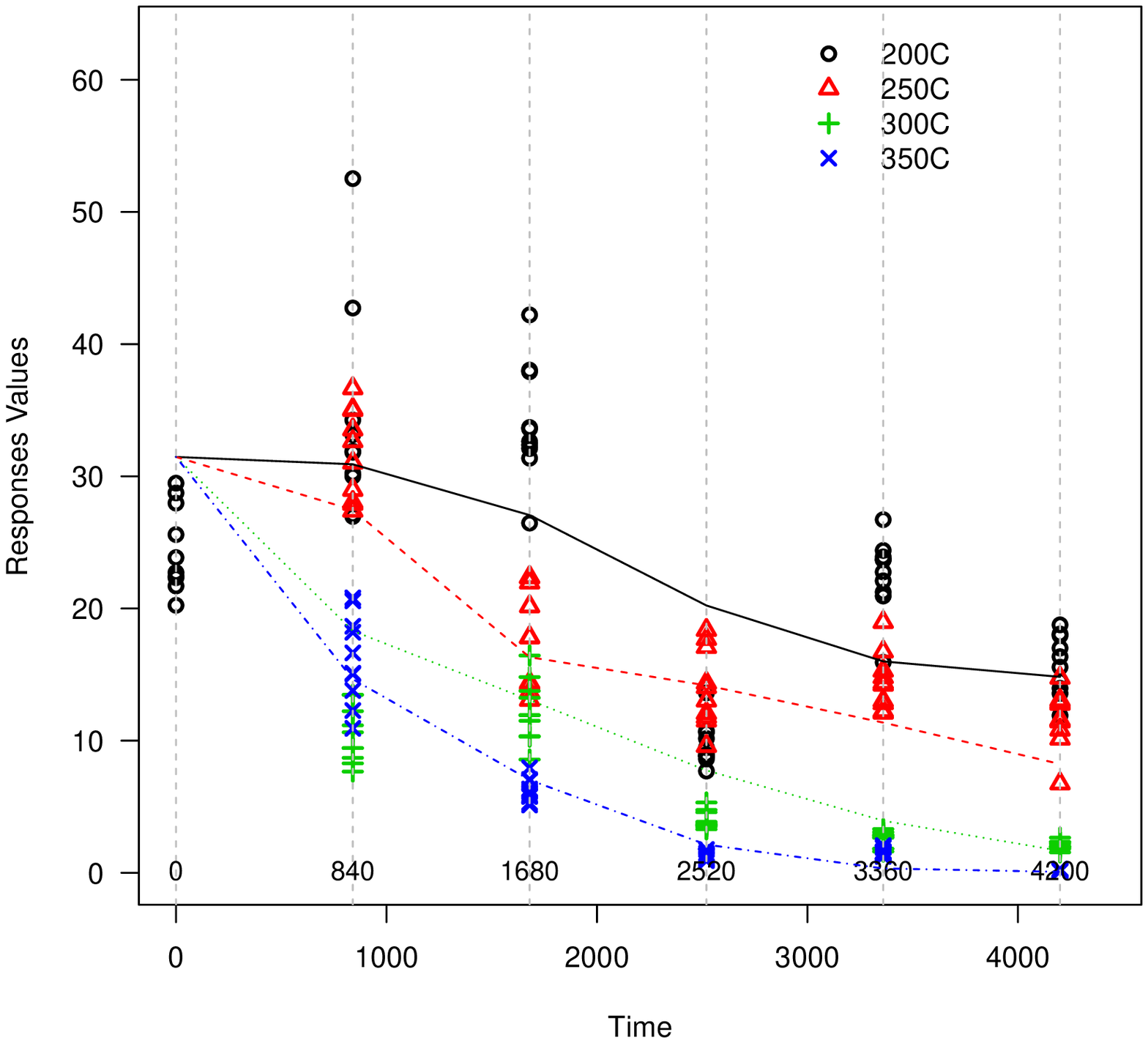}
\end{center}
\caption{Plots of fitted lines using the semiparametric method for the Seal Strength data, for the model without $\rho$. }
\label{plot: Seal Strength SP no cor}
\end{figure}

\begin{figure}
\begin{center}
\includegraphics[width=.8\textwidth]{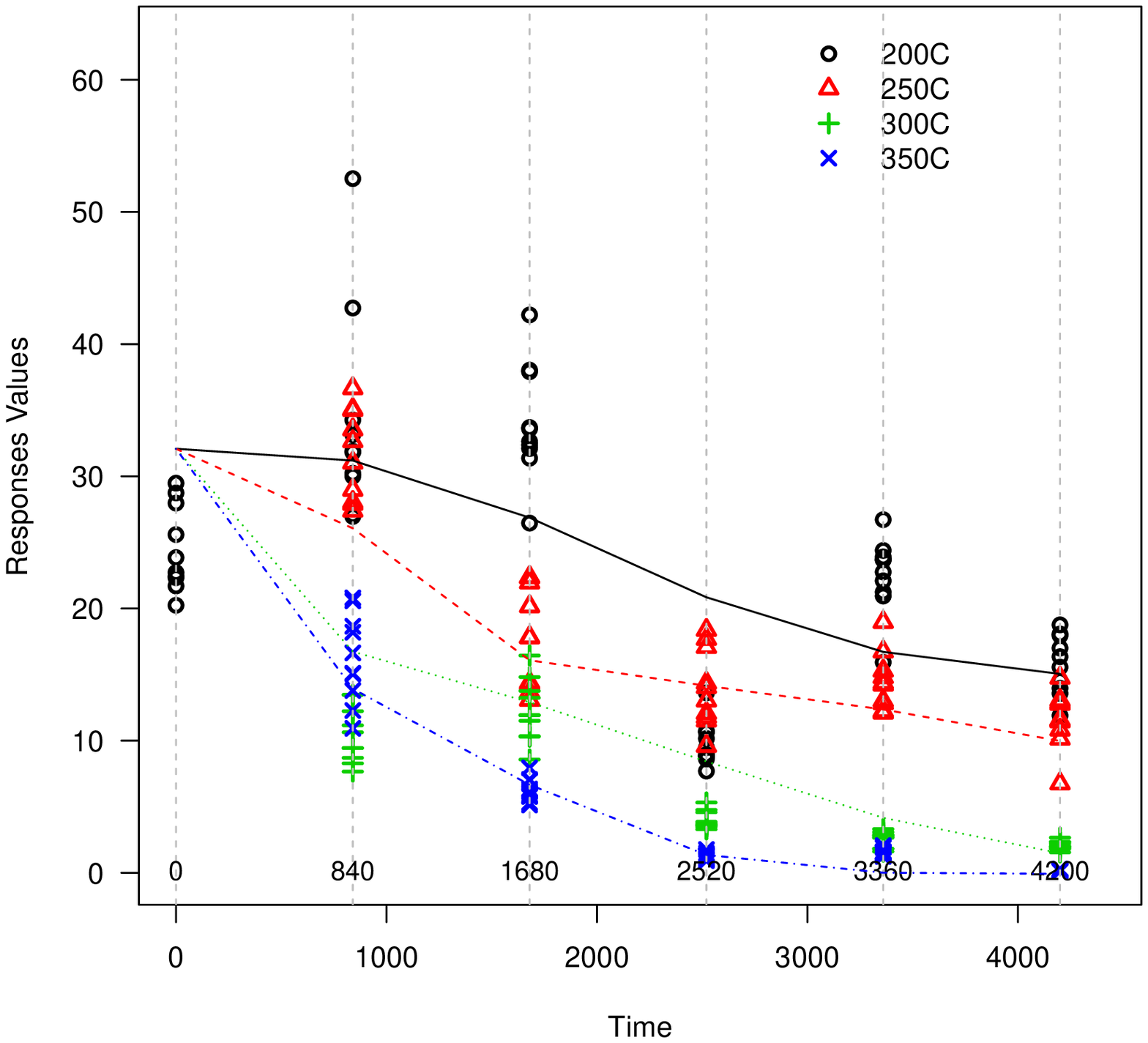}
\end{center}
\caption{Plots of fitted lines using the semiparametric method for the Seal Strength data, for the model with $\rho$.}\label{plot: Seal Strength SP cor}
\end{figure}

Similar to Section \ref{sec:R examples}, we illustrate the results from the least-squares, the maximum likelihood, and the semiparametric methods in Figures \ref{plot: Seal Strength LS}, \ref{plot: Seal Strength ML}, \ref{plot: Seal Strength SP no cor} and \ref{plot: Seal Strength SP cor}, respectively. Note that in Figures \ref{plot: Seal Strength SP no cor} and \ref{plot: Seal Strength SP cor}, we show the results for models without $\rho$ and with $\rho$, respectively.

\begin{figure}
\begin{center}
\includegraphics[width=.8\textwidth]{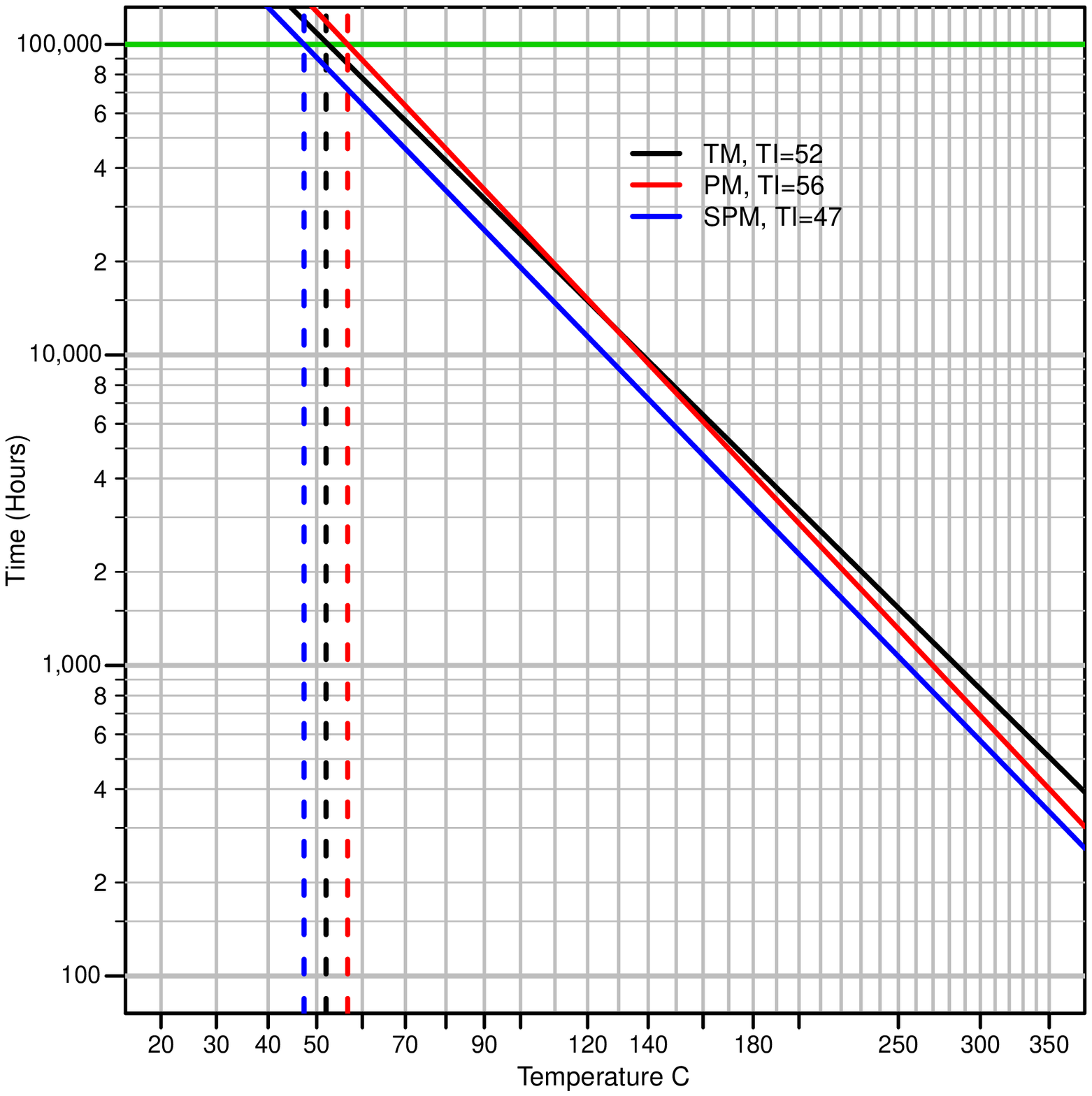}
\end{center}
\caption{Fitted temperature-time relationship lines for Seal Strength data using the traditional, maximum likelihood, and semiparametric methods. The failure threshold is 70\%.}
\label{plot: TI plot}
\end{figure}

In addition, users can specify the \emph{semi.control} argument in the \emph{SemiPara} fit option. The \emph{semi.control} contains a list of arguments that regards the \emph{SemiPara} option in the model. For example, whether or not to include a correlation \emph{$\rho$} in the model. When \emph{semi.control = list(cor = T)}, the model will fit the correlation model with $\rho$. Otherwise, when default value \emph{semi.control = list(cor = F)} or \emph{semi.control} is not specified, the no-correlation model will be fitted. Note that for the option \emph{SemiPara} in the function \emph{addt.fit}, including the correlation $\rho$ in the model may require more computing time, but potentially it will provide a better fit.

Here we compare the model results from the traditional method, the parametric method, and the semiparametric method for the Seal Strength data. In the results from \emph{summary}, TI estimates are 52$\degreeC$, 56$\degreeC$  and 47$\degreeC$, respectively. With $\beta_{0}$ and $\beta_{1}$ estimates, the TI plot is presented in Fig \ref{plot: TI plot}. The black line is the TI from the traditional model, the red line is the parametric model TI estimates, and the blue line stands for the results from the semiparametric method.

In the results from two methods, without and with the correlation $\rho$, $\hat\beta$ are 0.282 and 0.323, while TI estimates are 32.768 and 47.338, respectively. The differences come from the assumption of $\rho$ in the model. From the AICc value, the model with correlation provides a better fit to the data because it provides a smaller AICc value. The details of the model outputs are shown as follows.

\begin{verbatim}
>addt.fit.semi.no.cor<-addt.fit(Response~TimeH
+TempC,data=SealStrength,proc="SemiPara",
failure.threshold=70)

>addt.fit.semi.cor<-addt.fit(Response~TimeH
+TempC,data=SealStrength,proc="SemiPara",
failure.threshold=70, semi.control = list(cor=T))
\end{verbatim}

\begin{itemize}
\item Model without correlation $\rho$:
\end{itemize}

\begin{verbatim}
> summary(addt.fit.semi.no.cor)

 Semi-Parametric Approach:

Parameters Estimates:
betahat
  0.282

TI estimates:
 TI.semi    beta0    beta1
  32.768    0.362 1418.833

Model Evaluations:
Loglikelihood     AICC
     -639.206 1288.412

B-spline:
 Left Boundary    knots     knots     knots    knots
          0.00   268.60    527.17    840.00  1394.55
 Right Boundary
 4200.00
\end{verbatim}

\begin{itemize}
\item Model with correlation $\rho$:
\end{itemize}

\begin{verbatim}
> summary(addt.fit.semi.cor)

 Semi-Parametric Approach:

Parameters Estimates:
betahat     rho
  0.323   0.714

TI estimates:
 TI.semi    beta0    beta1
  47.338   -0.087 1630.282

Model Evaluations:
Loglikelihood     AICC
     -552.662  1117.323

B-spline:
 Left Boundary    knots     knots     knots     knots
          0.00   265.59    520.02    840.00   2483.29
 Right Boundary
 4200.00
\end{verbatim}

%%%%%%%%%%%%%%%%%%%%%%%%%%%%%%%%%%%%%%%%%%%%%%%%%%%%%%%%%%%%%%%%%%%%%%%%%%%%%%%%%%%%%%%%%%%%%%%%
\section{Concluding Remarks}\label{sec:concluding.remarks}
%%%%%%%%%%%%%%%%%%%%%%%%%%%%%%%%%%%%%%%%%%%%%%%%%%%%%%%%%%%%%%%%%%%%%%%%%%%%%%%%%%%%%%%%%%%%%%%%
In this chapter, we provide a comprehensive description with illustrations for the ADDT methods implemented in the \texttt{ADDT} package. Functions such as the \emph{addt.fit} and \emph{summary} are illustrated for the traditional method, the parametric method, and the semiparametric method. We also show  R examples using the Adhesive Bond B data and the Seal Strength data under various function options like \emph{proc} and \emph{semi.control}. Results from three different models are provided and visualized.  Users can consult the reference manual \cite{Raddt} for further details regarding the software package.

%%%%%%%%%%%%%%%%%%%%%%%%%%%%%%%%%%%%%%%%%%%%%%%%%%%%%%%%%%%%%%%%%%%%%%%%%%%%%%%%%%%%%%%%%%%%%%%%%%%%%%%%%%%%%%%%%
\section*{Acknowledgments}
%%%%%%%%%%%%%%%%%%%%%%%%%%%%%%%%%%%%%%%%%%%%%%%%%%%%%%%%%%%%%%%%%%%%%%%%%%%%%%%%%%%%%%%%%%%%%%%%%%%%%%%%%%%%%%%%%
The authors acknowledge Advanced Research Computing at Virginia Tech for providing computational resources. The work by Hong was partially supported by the National Science Foundation under Grant CMMI-1634867 to Virginia Tech.


%%%%%%%%%%%%%%%%%%%%%%%%%%%%%%%%%%%%%%%%%%%%%%%%%%%%%%%%%%%%%%%%%%%%%%%%%%%%%%%%%%%%%%%%%
\end{document}